\documentclass[12pt]{article}
\usepackage{amsfonts,graphicx,amssymb,amsmath,amsthm,epsfig}
\usepackage{float}
\allowdisplaybreaks
\begin{document}
\title{\bf Cosmography of Non-Interacting Ghost and Generalized
Ghost Dark Energy Models in $f(Q)$ Gravity}
\author{M. Sharif \thanks {msharif.math@pu.edu.pk}~and
\ Madiha Ajmal \thanks {madihaajmal222@gmail.com} \\
Department of Mathematics and Statistics, The University of Lahore,\\
1-KM Defence Road Lahore-54000, Pakistan.}

\date{}
\maketitle
\begin{abstract}
This paper aims to develop non-interacting ghost dark energy and
generalized ghost dark energy models within the framework of $f(Q)$
theory using the correspondence scheme. We use pressureless matter
and a power-law scale factor. The cosmic implications of the
resulting models are studied through the equation of state parameter
and the phase planes. We also check the stability of the
reconstructed models through the squared speed of sound parameter.
The equation of state parameter exhibits a phantom era, the
$(\omega_{D}-\omega^{\prime}_{D})$-plane indicates a freezing
region, while the $(r-s)$-plane corresponds to the Chaplygin gas
model for both models. It is also found that only the generalized
ghost dark energy model remains stable throughout cosmic evolution.
We conclude that our findings align well with current observational
data.
\end{abstract}

\section{Introduction}

Cosmology represents one of the most complex domains of astronomy
\cite{1}, focusing on the study of the cosmic structure as well as
its origin and evolutionary trajectory. It encompasses an
exploration of fundamental questions about the universe such as its
composition, behavior and ultimate fate. The general theory of
relativity (GR) established the basis for contemporary cosmology.
One of the most notable accomplishments in this field is the rapid
expansion of the cosmos which has been confirmed by various
cosmological observations such as Supernovae type-Ia \cite{3},
large-scale structure \cite{4} and cosmic microwave background
radiation \cite{7}. The accelerated expansion of cosmos has been
attributed to a mysterious form of energy known as dark energy (DE).
This enigmatic form of energy is evenly spread throughout the cosmos
with an ambiguous nature. The $\Lambda$ cold dark matter
($\Lambda$CDM) model depends on the cosmological constant and serves
as the fundamental framework for understanding DE. Despite its
consistency with observational data, the $\Lambda$CDM model faces
several challenges including issues like fine-tuning and the cosmic
coincidence problems \cite{7a}. To comprehend the nature of DE,
researchers employed two different methods. Firstly, by examining DE
models and secondly, through the use of modified gravity theories.

Every DE model contains many concealed and distinct characteristics,
leading to a challenging scenario to researchers. Most DE models
typically involve an additional degree of freedom to account for the
current state of the cosmos \cite{8}. This additional factor may
lead to inconsistency in the outcomes. For a satisfactory DE model,
it is crucial to solve this issue without depending on any new
degree of freedom or an extra parameter. A novel form of DE called
Veneziano ghost DE (GDE) has been suggested to achieve this purpose
\cite{9}. The energy density of the GDE model is represented as
\begin{equation}\label{1}
\rho_D=\alpha H,
\end{equation}
here, $\alpha$ denotes an arbitrary constant with dimensions of
$[energy]^{3}$. The GDE arises from the need to account for the
observed accelerated expansion of the universe without resorting to
the introduction of exotic form of matter or energy. By considering
such unconventional energy configurations, GDE models aim to explain
cosmic acceleration while remaining consistent with theoretical
principles and observational constraints.

The GDE presents a fascinating concept in cosmology. In simple
terms, the ghost field contribution to vacuum energy varies
depending on the type of spacetime. This contribution scales
proportionally with $\Lambda_{QCD}^3 H$, where $\Lambda_{QCD}$
represents the quantum chromodynamics (QCD) \cite{10} mass scale. In
the GDE model, the vacuum energy associated with the ghost field
behaves similar to a dynamic cosmological constant. Extensive
research has been delved into various aspects of GDE. This model
deals with certain issues effectively but it faces stability
challenges \cite{15}. It is found that the Veneziano QCD ghost
field's contribution to vacuum energy is not exactly of order $H$
but rather involves a subleading term $H^{2}$, known as the
generalized GDE (GGDE) model. The density of this GGDE is described
as
\begin{equation}\label{2}
\rho_D=\alpha H+\beta H^{2},
\end{equation}
where $\beta$ denotes another arbitrary constant with dimensions of
$[energy]^{2}$. This model is an exciting area in cosmology that
aims to uncover the mysterious nature of DE. This theoretical
framework extends beyond traditional models, incorporating ghost
fields to address shortcomings and offers fresh insights into the
rapid expansion of the cosmos. This model studies ghost-like fields
in the universe to better understand how the universe evolves and
its overall structure.

The current rapid expansion of the cosmos can be explored through
various models of DE. Researchers often use reconstruction
techniques in modified gravity theories to create DE models that
explain the universe acceleration. This process compares the energy
densities of the DE model and modified gravity theory. The
correspondence between these energy densities has been widely
explored in various DE models in the literature. Using the
correspondence scheme, we examine how these models work together to
enhance our understanding of cosmic acceleration. This method helps
to connect theoretical predictions with observational data, offering
a deep understanding of how DE and modified gravity contribute.
Saaidi et al. \cite{16} used a correspondence scheme to reconstruct
the GDE $f(R)$ model ($R$ represents the Ricci scalar) to examine
its stability and evolution through the analysis of cosmological
parameters. Sheykhi and Movahed \cite{17} examined the implications
of the same model in GR by observing the expansion of the cosmos
through limitations on the model parameter. Sadeghi et al. \cite{18}
discussed the interaction of this model by modifying both the
gravitational constant and the cosmological constant. They conducted
numerical computations to study the behavior of the universe through
the equation of state (EoS) and deceleration parameters. Fayaz et
al. \cite{19} explored this concept by studying $f(R)$ gravity to
analyze the universe evolution.

Chattopadhyay \cite{20} discussed two realistic $f(T)$ models ($T$
is the torsion scalar) and examined their stability using the
squared speed of sound and EoS parameter. Pasqua et al. \cite{21}
explored DE within the framework of $f(R,\mathbb{T})$ gravity
($\mathbb{T}$ is the trace of the energy-momentum tensor (EMT)),
incorporating a higher order of the Hubble parameter. Fayaz et al.
\cite{22} studied this concept within $f(R,\mathbb{T})$ gravity and
determined that their findings support the present state of the
cosmos. Sharif and Saba \cite{23} proposed a reconstruction of GDE
and GGDE within the framework of modified Gauss-Bonnet gravity. Liu
\cite{24} studied the dynamics of the scalar field to characterize
intricate quintessence cosmology through the ghost model of
interacting DE. Zarandi and Ebrahimi \cite{25} studied the cosmic
age problem in holographic DE and GGDE models. Sharif and Ibrar
\cite{25a} discussed the reconstructed GDE $f(Q,\mathbb{T})$ gravity
model ($Q$ is the non-metricity) for the non-interacting case.

The Levi-Civita (LC) connection, which is torsion-free but
compatible with metric, forms the foundation of GR. This was
replaced in flat spacetime with a metric-compatible affine
connection that includes torsion. In this alternative framework,
torsion was permitted to fully characterize gravity. This particular
concept was first suggested by Einstein referred to as the metric
teleparallel gravity \cite{26}. The symmetric teleparallel theory, a
recent addition to this family of theories, utilizes an affine
connection with zero curvature and torsion. This theory explores
gravity by emphasizing the non-metricity of spacetime \cite{27}. In
metric teleparallelism, the torsion scalar is derived from the
torsion tensor, whereas in the symmetric teleparallel framework, the
non-metricity scalar $Q$ is used for this purpose. By using the
Lagrangian $L=\sqrt{-g}T$ \cite{28} for the metric teleparallel
theory and $L=\sqrt{-g}Q$ \cite{29} for the symmetric teleparallel
theory, we can derive their field equations. Both theories are
equivalent to GR except for a boundary term. Consequently, both
metric and symmetric teleparallel theories address the same DE
issues as GR. To tackle this problem, new concepts related to
gravity, like $f(T)$ and $f(Q)$ have been put forward in their
specific frameworks. It is important to highlight that the
teleparallel theory differs from GR as the affine connection is not
linked to the metric tensor. Consequently, teleparallel theories are
classified within the broader category of metric-affine theories,
where both the metric and the connection serve as dynamic variables.
Notably, $f(T)$ and $f(Q)$ theories offer an advantage over $f(R)$
as they feature second-order field equations rather than
fourth-order ones. This distinction could offer a different
explanation for the observed acceleration of cosmic expansion
\cite{30}.

Researchers are highly interested in investigating non-Riemannian
geometry, particularly, the $f(Q)$ theory of gravity. Frusciante
\cite{39} proposed a particular model within this gravity framework
that showed foundational similarities to the $\Lambda$CDM model.
Ayuso et al. \cite{40} performed the comparison between the new
cosmologies and the $\Lambda$CDM setup and carried out the
statistical tests based on background observational data. Solanki et
al. \cite{41} investigated the impact of bulk viscosity in $f(Q)$
gravity to explore cosmic accelerated expansion. Albuquerque and
Frusciante \cite{42} investigated the evolution of linear
perturbations under the same theory. Sokoliuk et al. \cite{43}
described the evolution of the cosmos through Pantheon data sets in
this theory. Esposito et al. \cite{44} focused on employing the
reconstruction technique to investigate the precise isotropic and
anisotropic cosmological solutions in the same background. In a
recent paper \cite{45}, we have investigated the cosmography of GGDE
within the interacting scenario under the same gravity framework.
Gadbail and Sahoo \cite{46} studied three different models of this
to determine which one more accurately expresses theoretical
evolution of the $\Lambda$CDM model. Sharif et al. \cite{47}
investigated the concept of a cosmological bounce within the
framework of non-Riemannian geometry.

This paper examines the reconstruction of GDE and GGDE $f(Q)$ models
in the non-interacting scenario using a correspondence scheme. We
explore the evolution of the cosmos by analyzing the EoS, phase
planes and the squared speed of sound ($\nu_{s}^{2}$). The article
is structured as follows. Section \textbf{2} outlines the details of
this modified theory of gravity. We also discuss non-interacting DE
and DM in the FRW universe background. In sections \textbf{3} and
\textbf{4}, we explore a correspondence technique to reconstruct
both models in $f(Q)$ gravity. Finally, section \textbf{5} presents
a summary of our findings.

\section{Overview of $f(Q)$ Theory}

A connection which is both torsion-free as well as compatible with
the metric tensor is the LC connection \cite{48}. However, it is
possible to define two rank-3 tensors, torsion tensor
$(T_{\psi\gamma}^{\lambda})$ and non-metricity tensor
$(Q_{\gamma\psi\sigma})$, associated with the asymmetric tensor and
covariant derivative of the metric tensor, respectively, as
\begin{equation}\label{3}
T_{\psi\gamma}^{\lambda}=2\hat{\Gamma}^{\lambda}_{[\psi\gamma]},~~
Q_{\gamma\psi\sigma}=\nabla_{\sigma}g_{\gamma\psi}\neq 0.
\end{equation}
In this type of spacetime, the asymmetric tensor \cite{49} can be
represented as
\begin{equation}\label{4}
\hat{\Gamma}^{\lambda}_{\psi\gamma}={\Gamma}^{\lambda}_{\psi\gamma}
+\mathbb{C}^{\lambda}_{\;\psi\gamma}+\mathbb{L}^{\lambda}_{\;\psi\gamma},
\end{equation}
where, the LC connection is given as follows
\begin{equation}\label{5}
\Gamma^{\lambda}_{\psi\gamma}=\frac{1}{2}g^{\lambda\sigma}
(g_{\sigma\psi,\gamma}+g_{\sigma\gamma,\psi}-g_{\psi\gamma,\sigma}),
\end{equation}
the contortion tensor is expressed as
\begin{equation}\label{6}
\mathbb{C}^{\lambda}_{\;\psi\gamma}=\hat{\Gamma}^{\lambda}_{[\psi\gamma]}
+g^{\lambda\sigma}g_{\psi\kappa}\hat{\Gamma}^{\kappa}_{[\gamma\sigma]}
+g^{\lambda\sigma}g_{\gamma\kappa}\hat{\Gamma}^{\kappa}_{[\psi\sigma]}
\end{equation}
and the disformation tensor is given by
\begin{equation}\label{7}
\mathbb{L}^{\lambda}_{\;\psi\gamma}=\frac{1}{2}g^{\lambda\sigma}(Q_{\gamma\psi\sigma}
+Q_{\psi\gamma\sigma}-Q_{\lambda\psi\gamma}).
\end{equation}
Within the framework of symmetric connection, the LC connection can
be expressed in terms of the disformation tensor as follows
\begin{equation}\label{8}
\Gamma^{\lambda}_{\gamma\psi}=-\mathbb{L}^{\lambda}_{\;\gamma\psi}.
\end{equation}
In GR, it is widely acknowledged that the gravitational action can
be rewritten into a non-covariant form as
\begin{equation}\label{9}
S=\frac{1}{2k}\int g^{\gamma\psi}(\Gamma^{\mu}_{\sigma\gamma}
\Gamma^{\sigma}_{\psi\mu} -\Gamma^{\mu}_{\sigma\mu}
\Gamma^{\sigma}_{\gamma\psi})\sqrt{-g} \textrm{d}^ {4}x.
\end{equation}
Using Eq.\eqref{8}, the gravitational action can be represented as
\begin{equation}\label{10}
S=-\frac{1}{2k} \int g^{\gamma\psi}(\mathbb{L}^{\mu}_{~\sigma\gamma}
\mathbb{L}^{\sigma}_{~\psi\mu} - \mathbb{L}^{\mu}_{~\sigma\mu}
\mathbb{L}^{\sigma}_{~\gamma\psi}) \sqrt{-g} \textrm{d}^ {4}x.
\end{equation}
This is the action of the symmetric teleparallel gravity.

We investigate an extension of the symmetric teleparallel gravity
Lagrangian \eqref{10} as \cite{29}
\begin{equation}\label{11}
S=\int\bigg(\mathcal{L}_{\mathbf{m}}+\frac{1}{2k}f(Q)\bigg)
\sqrt{-g} \textrm{d}^{4}x,
\end{equation}
where, $g$ denotes determinant of the metric $g_{\psi\gamma}$,
$\mathcal{L}_{\mathbf{m}}$  represents the matter Lagrangian
density, and $f(Q)$ is a general function of $Q$. This function can
be described as
\begin{equation}\label{12}
Q=-g^{\gamma\psi}(\mathbb{L}^{\mu}_{~\nu\gamma}\mathbb{L}^{\nu}_{~\psi\mu}
-\mathbb{L}^{\mu}_{~\nu\mu}\mathbb{L}^{\nu}_{~\gamma\psi}).
\end{equation}
Utlizing Eq.\eqref{5} into Eq.\eqref{8}, it follows that
\begin{equation}\label{13}
\mathbb{L}^{\mu}_{\;\nu\varsigma}=-\frac{1}{2}g^{\mu\lambda}
(\nabla_{\varsigma}g_{\nu\lambda}+\nabla_{\nu}g_{\lambda\varsigma}
-\nabla_{\lambda}g_{\nu\varsigma}).
\end{equation}
The traces of the $Q$ tensor are
\begin{equation}\label{14}
Q_{\mu}=Q^{~\psi}_{\mu~\psi},\quad
\tilde{Q}_{\mu}=Q^{\psi}_{~\mu\psi}.
\end{equation}
The superpotential can be expressed as
\begin{equation}\label{15}
\mathbb{P}^{\mu\psi\gamma}=\frac{1}{4}\big[-Q^{\mu\psi\gamma}+Q^{\psi\mu\gamma}
+Q^{\gamma\mu\psi}+Q^{\psi\mu\gamma}-\tilde{Q}_{\mu}g^{\psi\gamma}
+Q^{\mu}g^{\psi\gamma}\big].
\end{equation}
As a result, the relation for $Q$ becomes \cite{45}
\begin{equation}\label{16}
Q=-Q_{\mu\gamma\psi}\mathbb{P}^{\mu\gamma\psi}=-\frac{1}{4}(-Q^{\mu\psi\rho}Q_{\mu\psi\rho}
+2Q^{\mu\psi\rho}Q_{\rho\mu\psi}-2Q^{\rho}\tilde{Q}_{\rho}+Q^{\rho}Q_{\rho}).
\end{equation}
The field equations of $f(Q)$ gravity turn out to be
\begin{equation}\label{17}
\frac{-2}{\sqrt{-g}}\nabla_{\gamma}(f_{Q}\sqrt{-g}
P^{\mu}_{~\gamma\psi})-\frac{1}{2}f g_{\gamma\psi}-f_{Q}
(P_{\gamma\mu\nu}Q_{\psi}^{~\mu\nu}-2Q^{\mu\nu}_{~~~\gamma}
P_{\mu\nu\psi})=k^{2} \mathbb{T}_{\gamma\psi},
\end{equation}
where $f_{Q}=\frac{\partial f(Q)}{\partial Q}$.

The line element describing a spatially homogeneous and isotropic
model of the cosmos is expressed as
\begin{equation}\label{20}
\textrm{d}s^{2}=a^{2}(t)(\textrm{d}x^{2}+\textrm{d}y^{2}
+\textrm{d}z^{2})-\textrm{d}t^{2},
\end{equation}
where the scale factor is represented by $a(t)$. The total EMT for
DE and DM is defined as follows
\begin{equation}\label{21}
\hat{\mathbb{T}}_{\psi\gamma}=\mathbb{T}_{\psi\gamma}+\tilde{\mathbb{T}}_{\psi\gamma},
\end{equation}
where the EMT for pressureless DM is
\begin{equation}\label{22}
\mathbb{T}_{\psi\gamma}=(\rho_{\mathbf{m}})u_{\psi}u_{\gamma}
\end{equation}
and for DE is
\begin{equation}\label{23}
\tilde{\mathbb{T}}_{\psi\gamma}=(\rho_{D}+p_{D})u_{\psi}u_{\gamma}+p_{D}g_{\psi\gamma}.
\end{equation}
Here, $\rho_{\mathbf{m}}$ and $\rho_{D}$ denote the energy densities
of DM and DE, respectively, while $p_{D}$ stands for the pressure of
DE and $u_{\gamma}$ represents the four velocity field. The modified
Friedman equations for $f(Q)$ gravity are expressed as
\begin{equation}\label{18}
3H^{2}=\rho_D +\rho_{\mathbf{m}},\quad 2\dot{H}+3H^{2}=p_D
+p_{\mathbf{m}},
\end{equation}
where
\begin{eqnarray}\label{19}
\rho_D&=&-6H^{2}f_{Q}+\frac{f}{2},\\\label{19a}
p_D&=&2Hf_{QQ}+\frac{f}{2}+6H^{2}f_{Q}+2f_{Q}\dot{H},
\end{eqnarray}
where, dot signifies derivative with respect to $t$.

The continuity equation for the non-interacting DM and DE are
\begin{eqnarray}\label{24}
\dot{\rho}_{\mathbf{m}}+3H(\rho_{\mathbf{m}})&=&0,\\\label{25}
\dot{\rho}_{D}+3H(\rho_{D}+p_{D})&=&0.
\end{eqnarray}
To find the analytical solution, we assume the scale factor in the
form of a power law as
\begin{equation}\label{26}
a(t)=a_{0}t^{m},
\end{equation}
where $m$ and $a_0$ are arbitrary constants, $a_0=1$ is taken as the
current value. In terms of cosmic time $t$, we can utilize this
correlation to express the values of $H$, its derivative and the
non-metricity scalar
\begin{equation}\label{27}
H=\frac{\dot{a}}{a}=\frac{m}{t}, \quad\dot{H}=-\frac{m}{t^{2}},
\quad Q=6H^{2}=6 \frac{m^2}{t^2}.
\end{equation}
Integrating Eq.\eqref{24}, we have
\begin{equation}\label{28}
\rho_{\mathbf{m}}=\rho_{0}(a)^{-3}=\rho_{0}(t^m)^{-3},
\end{equation}
where $\rho_{0}$ is an integration constant.

\section{Reconstruction of GDE $f(Q)$ Model}

In this section, we reconstruct the GDE $f(Q)$ model by equating the
corresponding densities. Applying Eqs.\eqref{1} and \eqref{19}, it
follows that
\begin{equation}\label{29}
\frac{f}{2}-6H^{2}f_{Q}=\alpha H.
\end{equation}
This is the first-order linear differential equation involving the
variable $Q$ and its solution is
\begin{equation}\label{30}
f(Q)=\frac{1}{6} \sqrt{Q} \bigg(6 c-\sqrt{6} \alpha\ln (Q)\bigg),
\end{equation}
where $c$ is the integration constant. In terms of $t$, this model
can be derived by inserting \eqref{27} into \eqref{30} as
\begin{equation}\label{31}
f(Q)=\frac{\sqrt{\frac{m^2}{t^2}} \bigg(6 c-\sqrt{6} \alpha  \ln
\big(\frac{6 m^2}{t^2}\big)\bigg)}{\sqrt{6}}.
\end{equation}
Throughout the graphical analysis, we assume a constant value of
$c=0.5$. Figure \textbf{1} illustrates that the reconstructed GDE
model increases with respect to $t$ while maintaining a positive
value, which indicates a rapid expansion. Inserting Eq.\eqref{30}
into \eqref{19} and \eqref{19a}, we obtain $\rho_D$ and $p_D$ as
\begin{eqnarray}\label{32}
\rho_D&=&\frac{\alpha  \sqrt{Q}}{\sqrt{6}}+c,\\\nonumber
p_D&=&\frac{1}{12 Q^{\frac{3}{2}}}\bigg[c Q^2-4 \dot{H} Q
\big(\sqrt{6} \alpha -3 c\big)+6 \big(H^2 Q \big(6 c-2 \sqrt{6}
\alpha \big)-c H\big)\\\label{33} &-&\sqrt{6} \alpha  \ln (Q) \big(Q
(2 \dot{H}+Q)+6 H^2 Q-H\big)\bigg].
\end{eqnarray}
Using Eq.\eqref{27} into the above equations, we can express them in
power-law form as
\begin{figure}\center
\epsfig{file=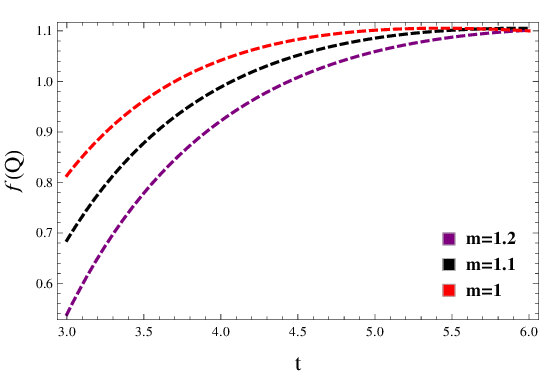,width=0.5\linewidth} \caption{The graph
illustrates the relationship between $f(Q)$ and $t$.}
\end{figure}
\begin{eqnarray}\label{1aa}
\rho_D&=& \alpha  \sqrt{\frac{m^2}{t^2}}+c,\\\nonumber
p_D&=&\frac{1}{72 \sqrt{6} t^4
\big(\frac{m^2}{t^2}\big)^{3/2}}\bigg[m\bigg(\sqrt{6} \alpha
\big(-72 m^3+12 m^2+t^3\big) \ln \bigg(\frac{6
m^2}{t^2}\bigg)\\\label{1bb} &-&6 \big(12 m^3 \big(\sqrt{6} \alpha
-6 c\big)-4 m^2 \big(\sqrt{6} \alpha -3 c\big)+c
t^3\big)\bigg)\bigg].
\end{eqnarray}
\begin{figure}
\epsfig{file=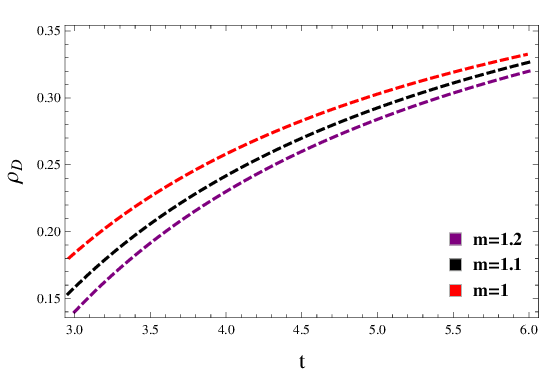,width=.5\linewidth}
\epsfig{file=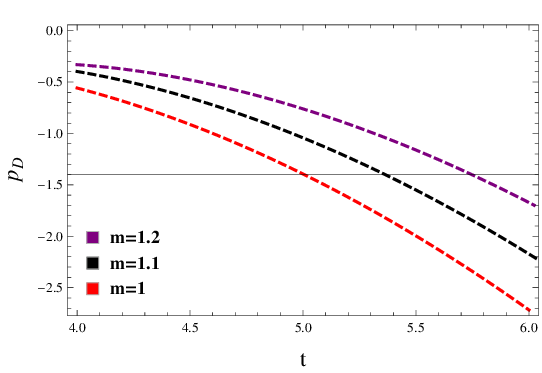,width=.5\linewidth}\caption{Graphs show $\rho_D$
and $p_D$ against $t$}
\end{figure}

Figure \textbf{2} illustrates the behavior of $\rho_D$ and $p_D$. It
is seen that the energy density $\rho_D$ remains positive and
exhibits an increasing pattern, consistent with the characteristics
of DE, while $p_D$ demonstrates negative behavior with a decreasing
trend. This positive energy density is not only stable but also
increases over time, aligning with the theoretical expectations for
DE models, where the energy density plays a crucial role in driving
the accelerated expansion of the universe. On the other hand, this
negative pressure is necessary to explain the repulsive
gravitational effect responsible for the universe's accelerated
expansion. The decreasing trend of $p_D$ in the graph highlights
this behavior, reinforcing the idea that DE dominates over other
forms of energy in the universe at late times, causing the
accelerated expansion.

Now, we explore the evolution of the cosmos by conducting
cosmographic analysis on the EoS parameter and phase planes for the
GDE $f(Q)$ model. We also discuss stability of this model by
examining the squared sound speed. The EoS is defined as
$\omega_{D}=\frac{p_{D}}{\rho_{D}}$, which is essential for
determining the dynamics of expansion. Specific values of
$\omega_{D}$ correspond to distinct types of energy contributions to
the universe. For example, $\omega_{D}= 0$ indicates a
matter-dominated region, typically associated with dust, while
$\omega_{D} = \frac{1}{3}$ corresponds to radiation and $\omega_{D}
= 1$ signifies stiff matter. In the context of DE, various ranges of
$\omega_{D}$ characterize different phases of expansion:
$\omega_{D}= -1$ denotes a cosmological constant representing vacuum
energy, $\omega_{D}<-1$ indicates the presence of phantom energy and
$-1<\omega_{D}<-\frac{1}{3}$ represents quintessence. By analyzing
the EoS parameter, researchers can gain insights into the dynamics
of cosmic expansion and the nature of the universe energy
components. This can be evaluated as follows
\begin{figure}\center
\epsfig{file=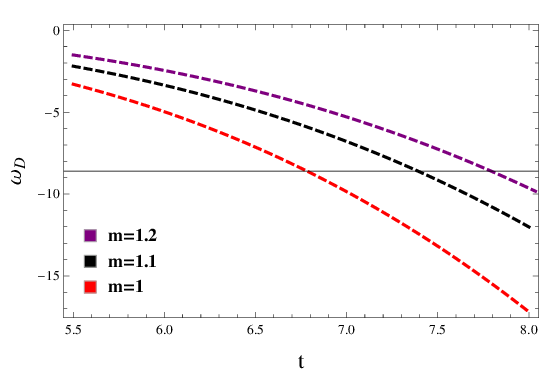,width=.6\linewidth}\caption{Plot shows the
relationship between $\omega_{D}$ with $t$.}
\end{figure}
\begin{eqnarray}\nonumber
\omega_{D}&=&\frac{1}{72 \sqrt{6} \bigg(\alpha  m^3 \xi +m
\sqrt{\frac{m^2}{t^2}} t^{2-3 m}\bigg)}\bigg[\xi  \bigg(6 \bigg(c
\big(72 m^3-12 m^2-t^3\big)+4\sqrt{6}\\\label{1cc} &\times& \alpha
(1-3 m) m^2\bigg)+\sqrt{6} \alpha \big(-72 m^3+12 m^2+t^3\big) \ln
\bigg(\frac{6 m^2}{t^2}\bigg)\bigg)\bigg].
\end{eqnarray}
Figure \textbf{3} demonstrates the behavior of the EoS in GDE $f(Q)$
gravity for three distinct values of $m=1.2,~ 1.1,~ 1$, which
represents $\omega_{D}<-1$, indicating the presence of phantom field
DE. This suggests that the universe is undergoing accelerated
expansion at a rate faster than predicted by conventional models.
Such expansion can significantly affect how the universe evolves and
may influence its ultimate fate. By analyzing the EoS for these
values, it reveals the complex relationship between parameters in
$f(Q)$ gravity and DE, providing valuable insights into how modified
gravity theories enhance our understanding of cosmic acceleration.

The ($\omega_{D}-\omega^{\prime}_{D}$)-plane \cite{50} is used to
study the dynamics and properties of DE models. Here
$\omega^{\prime}_{D}$ represents its derivative with respect to $Q$.
We classify the DE models into two distinct categories, i.e.,
thawing and freezing regions. When $\omega_{D}<0$ and
$\omega^{\prime}_{D}>0$, we have thawing region, while in the
freezing region, $\omega_{D}<0$ and $\omega^{\prime}_{D}<0$. Thus we
have
\begin{eqnarray}\nonumber
\omega^{\prime}_{D}&=&\frac{1}{864 \sqrt{6} m^3
\sqrt{\frac{m^2}{t^2}} \bigg(\alpha  \xi  \sqrt{\frac{m^2}{t^2}}
t^{3 m}+1\bigg)^2}\bigg[\xi  t^{3 m} \bigg(2 \bigg(c \bigg( \bigg(2
\alpha  \xi  \sqrt{\frac{m^2}{t^2}} t^{3 m}+1\bigg)\\\nonumber
&\times&36 m^2+3 t^3 \bigg(4 \alpha  \xi \sqrt{\frac{m^2}{t^2}} t^{3
m}+3\bigg)-216 \alpha  m^3 \xi \sqrt{\frac{m^2}{t^2}} t^{3 m}\bigg)-
\alpha  \bigg(36 m^3\\\nonumber &+&12 \alpha  \xi t^{3 m+2}
\bigg(\frac{m^2}{t^2}\bigg)^{3/2}-t^3 \bigg(\alpha  \xi
\sqrt{\frac{m^2}{t^2}} t^{3 m}+1\bigg)\bigg)\sqrt{6}\bigg)+ \alpha
\ln \bigg(\frac{6 m^2}{t^2}\bigg) \\\nonumber
&\times&\sqrt{6}\bigg(-12 m^2 \bigg(2 \alpha  \xi
\sqrt{\frac{m^2}{t^2}} t^{3 m}+1\bigg)-t^3 \bigg(4 \alpha  \xi
\sqrt{\frac{m^2}{t^2}} t^{3 m}+3\bigg)\\\label{1dd} &+&72 \alpha m^3
\xi \sqrt{\frac{m^2}{t^2}} t^{3 m}\bigg)\bigg)\bigg].
\end{eqnarray}
Figure \textbf{4} illustrates that
$\omega_{D}<0,~\omega^{\prime}_{D}<0$ for various values of $m$,
indicating the existence of the freezing region. This suggests that
the acceleration of the cosmic expansion is increasing more
significantly in this model.
\begin{figure}\center
\epsfig{file=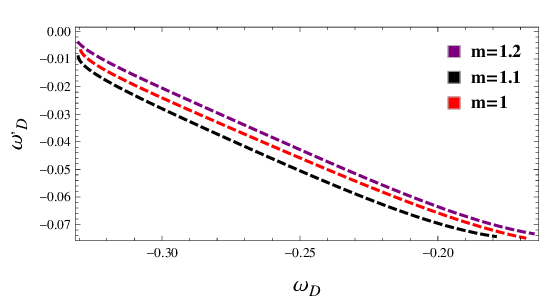,width=.6\linewidth}\caption{Graph of
$\omega^{\prime}_{D}$ against $\omega_{D}$.}
\end{figure}
\begin{figure}\center
\epsfig{file=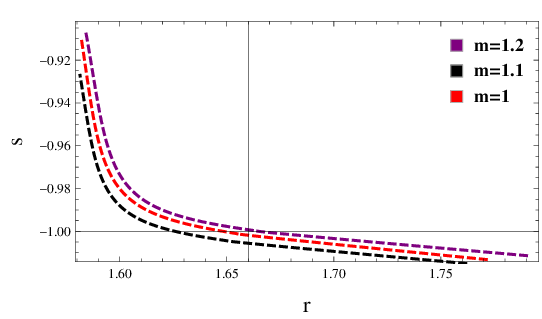,width=.6\linewidth}\caption{Graph depicts $r$
against $s$.}
\end{figure}

Statefinder parameters \cite{51} are cosmological diagnostic tools
used to characterize the expansion dynamics of the cosmos and
distinguish between different DE models. One can identify specific
trajectories that correspond to different DE models. Statefinder
parameters thus provide valuable insights into the nature and
behavior of DE, which help in comparing theoretical predictions with
observational data. These can be described as \cite{51}
\begin{equation}\nonumber
r=\frac{\dddot{a}}{aH^{3}}, \quad s=\frac{r-1}{3(q-\frac{1}{2})},
\end{equation}
Trajectories within the range $(r<1)$ and $(s>0)$ correspond to
phantom and quintessence DE eras, while trajectories with $(r>1)$
and $(s<0)$ indicate the Chaplygin gas model. The expressions for
the $(r-s)$ parameters are given in Appendix \textbf{A}. Figure
\textbf{5} indicates the Chaplygin gas model for three different
values of $m$. This figure highlights how varying $m$ affects the
behavior of the Chaplygin gas model, which is characterized by a
unified description of DE and DM. The results presented in this
figure demonstrate the versatility of the Chaplygin gas model in
explaining the accelerated expansion of the universe while
simultaneously offering insights into the interactions between DE
and modified gravity in the $f(Q)$ framework. The squared speed of
sound plays a crucial role to analyze the stability of different
models. In the DE models, when $\nu_{s}^{2}$ is positive it
characterizes the stability while a negative $\nu_{s}^{2}$ leads to
instability. The corresponding $\nu_{s}^{2}$ is given as
\begin{eqnarray}\nonumber
\nu_{s}^{2}&=&\frac{1}{5184 \sqrt{6} \alpha m^6 \bigg(\alpha  \xi
\sqrt{\frac{m^2}{t^2}} t^{3 m}+1\bigg)^2}\bigg[\xi
\sqrt{\frac{m^2}{t^2}} t^{3 m} \bigg(6 \bigg(12 m^3 \big(\sqrt{6}
\alpha -6 c\big)\\\nonumber&-& 4 m^2 \big(\sqrt{6} \alpha -3
c\big)+c t^3\bigg)+\sqrt{6} \alpha \big(72 m^3-12 m^2-t^3\big) \ln
\bigg(\frac{6 m^2}{t^2}\bigg)\bigg) \\\nonumber&\times&\bigg(-24
\alpha  (3 m-1) \xi \big(\alpha -\sqrt{6} c\big) t^{3 m+2}
\bigg(\frac{m^2}{t^2}\bigg)^{\frac{3}{2}}-2 \alpha  \xi \big(\alpha
+2 \sqrt{6} c\big) t^{3 m+3}
\\\nonumber&\times&\sqrt{\frac{m^2}{t^2}}-12 \sqrt{6} c m^2-t^3
\big(2 \alpha +3 \sqrt{6} c\big)-\alpha  \ln \bigg(\frac{6
m^2}{t^2}\bigg) \bigg(1- \bigg(2 \alpha  \xi t^{3
m}\\\label{1ee}&\times&\sqrt{\frac{m^2}{t^2}}\bigg)12 m^2-t^3
\bigg(4 \alpha  \xi \sqrt{\frac{m^2}{t^2}} t^{3 m}+3\bigg)+72 \alpha
m^3 \xi \sqrt{\frac{m^2}{t^2}} t^{3 m}\bigg)\bigg)\bigg].
\end{eqnarray}
\begin{figure}\center
\epsfig{file=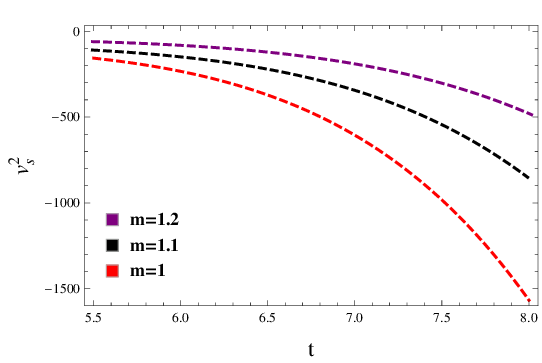,width=.6\linewidth}\caption{Graph of $\nu_s^{2}$
with $t$.}
\end{figure}

Figure \textbf{6} illustrates a consistently decreasing trend of
$\nu_{s}^{2}$, indicating the instability of the GDE model for
various values of $m$. This downward trend is important because it
indicates possible instability in the GDE framework. When
$\nu_{s}^{2}$ gets very low or negative, it means the model might
not behave as expected, which can affect its ability to explain how
the universe evolves. Understanding a stable DE model is crucial for
studying the universe's expansion. The results in this figure
highlight the challenges faced by the GDE model, suggesting the need
for further research on how different values of $m$ impact
stability. It is essential to explore changes to DE theories to
ensure they provide reliable descriptions of cosmic acceleration.

\section{Reconstruction of GGDE $f(Q)$ Model}
\begin{figure}\center
\epsfig{file=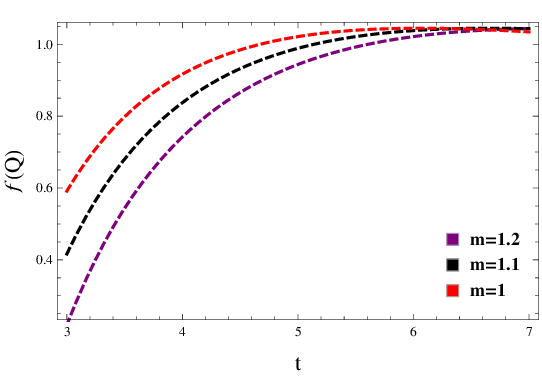,width=0.5\linewidth} \caption{Graphical
representation of $f(Q)$ versus $t$.}
\end{figure}

In this section, we reconstruct the GGDE $f(Q)$ model by equating
the corresponding densities equal to each other. Applying
Eqs.\eqref{2} and \eqref{19}, it is demonstrated that
\begin{equation}\label{34}
\frac{f}{2}-6H^{2}f_{Q}=\alpha H+\beta H^{2}.
\end{equation}
Its solution is given by
\begin{equation}\label{35}
f(Q)=-\frac{1}{6} \sqrt{Q} \bigg(-6 c+\sqrt{6} \alpha  \ln (Q)+2
\beta  \sqrt{Q}\bigg),
\end{equation}
and in terms of $t$, we have
\begin{equation}\label{36}
f(Q)=\sqrt{6} c \sqrt{\frac{m^2}{t^2}}-\alpha \sqrt{\frac{m^2}{t^2}}
\ln \bigg(\frac{6 m^2}{t^2}\bigg)-\frac{2 \beta  m^2}{t^2}.
\end{equation}
Figure \textbf{7} demonstrates that the reconstructed GGDE model
exhibits an increasing behavior with time,  while maintaining a
consistently positive value. This observation implies that the GGDE
model indicates an accelerated expansion of the universe. Using
Eq.\eqref{35} in \eqref{19} and \eqref{19a}, we have
\begin{figure}
\epsfig{file=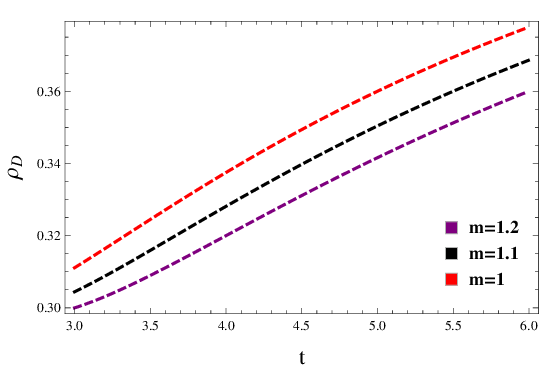,width=.5\linewidth}
\epsfig{file=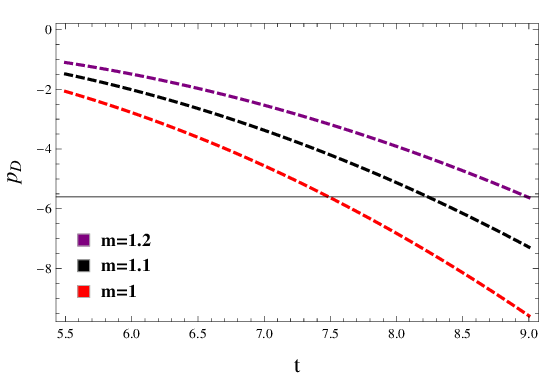,width=.5\linewidth}\caption{Plots of $\rho_D$
and $p_D$ versus $t$.}
\end{figure}
\begin{eqnarray}\label{37}
\rho_D&=&\frac{1}{6} \bigg(\sqrt{6} \alpha  \sqrt{Q}+\beta
Q\bigg)+c,\\\nonumber p_D&=&\frac{1}{12 Q^{\frac{3}{2}}}\bigg[-2
\big(3 c H-3 c Q^2+\big(\sqrt{6} \alpha -3 c+2 \beta
\sqrt{Q}\big)\big(2 \dot{H} Q +6H^2Q\big)
\\\label{38}&+&\beta Q^{\frac{5}{2}}\big)-\sqrt{6} \alpha
\ln Q \big(Q (2 \dot{H}+Q)+6 H^2 Q-H\big)\bigg].
\end{eqnarray}
Substituting Eq.\eqref{27} into the above equations, we can express
them in power-law form as
\begin{eqnarray}\label{2aa}
\rho_D&=&\alpha  \sqrt{\frac{m^2}{t^2}}+\frac{\beta
m^2}{t^2}+c,\\\nonumber p_D&=&\frac{1}{72 \sqrt{6} t^4
\big(\frac{m^2}{t^2}\big)^{\frac{3}{2}}}\bigg[m \bigg(\sqrt{6}
\alpha \big(12 m^2-72 m^3+t^3\big) \ln \bigg(\frac{6
m^2}{t^2}\bigg)-6 \bigg( c t^3\\\label{2bb}&+&\bigg(\sqrt{6} \alpha
-3 c+2 \sqrt{6} \beta \sqrt{\frac{m^2}{t^2}}\bigg)(12 m^3-4
m^2)\bigg)\bigg)\bigg].
\end{eqnarray}
Figure \textbf{8} shows the decreasing trend of $p_D$, while
$\rho_D$ maintains positivity and increases steadily. These patterns
align with the expected behavior of DE contributing to the
understanding of cosmic acceleration. The corresponding EoS
parameter takes the form
\begin{eqnarray}\nonumber
\omega_{D}&=&-\bigg\{t^{3 m} \bigg(6 \bigg(-4 m^2 \bigg(\sqrt{6}
\alpha -3 c+2 \sqrt{6} \beta  \sqrt{\frac{m^2}{t^2}}\bigg)+12 m^3
\bigg(\sqrt{6} \alpha -6 c\\\nonumber&+&3 \sqrt{6} \beta
\sqrt{\frac{m^2}{t^2}}\bigg)+c t^3\bigg)+\sqrt{6} \alpha \big(72
m^3-12 m^2-t^3\big) \ln \bigg(\frac{6
m^2}{t^2}\bigg)\bigg)\bigg\}\bigg\{72 \\\label{2cc}&\times&\sqrt{6}
m \sqrt{\frac{m^2}{t^2}} \bigg(\beta  m^2 t^{3 m}+\alpha
\sqrt{\frac{m^2}{t^2}} t^{3 m+2}+\xi t^2\bigg)\bigg\}^{-1}.
\end{eqnarray}
\begin{figure}\center
\epsfig{file=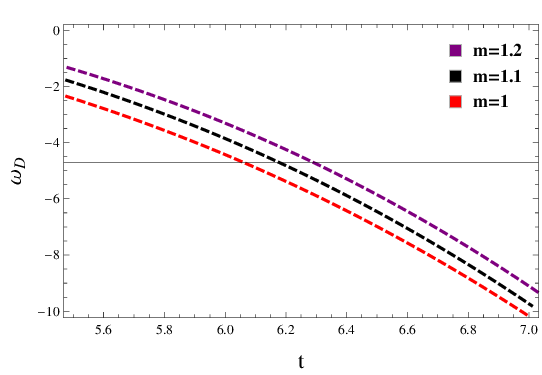,width=.6\linewidth}\caption{Graph of
$\omega_{D}$ with $t$.}
\end{figure}

Figure \textbf{9} shows the presence of a phantom epoch during both
the current and late-time cosmic evolution for various values of
$m$. The significance of a phantom epoch lies in its implications
for the dynamics of the universe's expansion. During this phase, the
energy density associated with DE increases, leading to an
accelerated expansion rate that exceeds predictions made by standard
cosmological models. This behavior raises the possibility of a
future cosmic event where the universe's expansion accelerates to a
point that could disrupt the fabric of spacetime. Overall, this
figure highlights the complex behavior of DE and underscores the
importance of exploring how modifications to existing models can
improve our understanding of cosmic evolution.

The expression of $\omega^{\prime}_{D}$ is given as
\begin{eqnarray}\nonumber
\omega^{\prime}_{D}&=&\bigg\{\sqrt{\frac{m^2}{t^2}} t^{3 m+4}
\bigg(\frac{1}{m^2}\bigg[2 \sqrt{\frac{m^2}{t^2}} \bigg(\xi  t^7
\big(\sqrt{6} \alpha +9 c\big) \sqrt{\frac{m^2}{t^2}}+m^2 t^4
\bigg(36 c \xi \sqrt{\frac{m^2}{t^2}}\\\nonumber&+&t^{3 m+1}
\bigg(\sqrt{6} \alpha ^2+12 \alpha c+15 \beta  c
\sqrt{\frac{m^2}{t^2}}+\sqrt{6} \alpha  \beta
\sqrt{\frac{m^2}{t^2}}\bigg)\bigg)+72 m^5 t^2 \bigg(t^{3
m}\\\nonumber&\times& \bigg(-3 \alpha  c-6 \beta  c
\sqrt{\frac{m^2}{t^2}}+\sqrt{6} \alpha \beta
\sqrt{\frac{m^2}{t^2}}\bigg)-\sqrt{6} \beta  \xi \bigg)-12 m^4 t^{3
m+2} \bigg(\sqrt{6} \alpha ^2\\\nonumber&-&6 \alpha  c-9 \beta  c
\sqrt{\frac{m^2}{t^2}}+4 \sqrt{6} \alpha  \beta
\sqrt{\frac{m^2}{t^2}}\bigg)+144 \sqrt{6} \beta ^2 m^7 t^{3 m}-48
\sqrt{6} \beta ^2 m^6 t^{3 m}\\\nonumber&-&36 \sqrt{6} \alpha  m \xi
t^6 \bigg(\frac{m^2}{t^2}\bigg)^{3/2}\bigg)\bigg]+\sqrt{6} \alpha
\ln \bigg(\frac{6 m^2}{t^2}\bigg) \bigg(144 \beta  m^5 t^{3 m}-36
\beta m^4 t^{3 m}\\\nonumber&-&m^2 t^2 \bigg(24 \alpha
\sqrt{\frac{m^2}{t^2}} t^{3 m}+5 \beta  t^{3 m+1}+12 \xi \bigg)-t^5
\bigg(4 \alpha \sqrt{\frac{m^2}{t^2}} t^{3 m}+3 \xi
\bigg)\\\label{2dd}&+&\frac{\alpha  m^5 t^{3
m}}{\sqrt{\frac{m^2}{t^2}}}\bigg)\bigg)\bigg\}\bigg\{864 \sqrt{6}
m^5 \bigg(\beta  m^2 t^{3 m}+\alpha  \sqrt{\frac{m^2}{t^2}} t^{3
m+2}+\xi  t^2\bigg)^2\bigg\}^{-1}.
\end{eqnarray}
Figure \textbf{10} demonstrates the cosmic trajectories on the
$\omega_{D}-\omega^{\prime}_{D}$ plane for certain values of $m$,
illustrating the freezing region of the universe. This plane
represents the current cosmic expansion paradigm, where the freezing
region suggests a phase of accelerated expansion compared to the
thawing region. This behavior is crucial as it suggests that the
universe may experience a stable acceleration in its expansion over
time. By analyzing the trajectories for various values of $m$, this
figure provides insights into how modifications to the GGDE model
influence cosmic evolution. Each trajectory illustrates how changes
in $m$ can affect the relationship between $\omega_{D}$ and its
derivative $\omega^{\prime}_{D}$, revealing patterns that
characterize the interaction between DE and cosmic expansion.
Overall, this figure serves as a vital tool for exploring the
dynamics of DE and its implications for the evolution of the
universe, emphasizing the significance of the freezing region in
understanding cosmic acceleration.
\begin{figure}\center
\epsfig{file=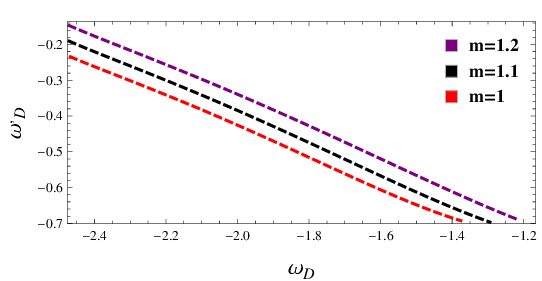,width=.6\linewidth}\caption{Plot of
$\omega^{\prime}_{D}$ versus $\omega_{D}$.}
\end{figure}
\begin{figure}\center
\epsfig{file=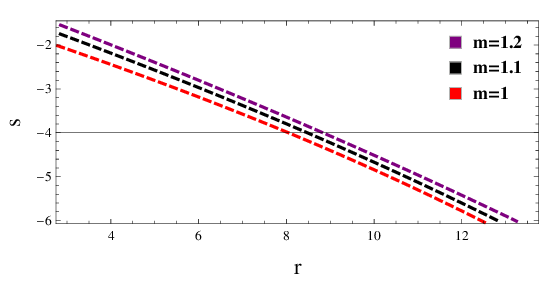,width=.6\linewidth}\caption{Plot of $r$ versus
$s$.}
\end{figure}

The $(r-s)$-plane is given in Appendix \textbf{B}. Figure
\textbf{11} depicts the graphical examination of the $r-s$ plane.
This shows that $r>1$ and $s<0$, indicating the presence of the
Chaplygin gas model. The expression for $\nu_{s}^{2}$ is provided as
follows
\begin{eqnarray}\nonumber
\nu_{s}^{2}&=&-\bigg\{t^{3 m} \bigg(6 \bigg(c t^3-4 m^2
\bigg(\sqrt{6} \alpha -3 c+2 \sqrt{6} \beta
\sqrt{\frac{m^2}{t^2}}\bigg)+12 m^3 \bigg(\sqrt{6} \alpha -6
c\\\nonumber &+&3 \sqrt{6} \beta
\sqrt{\frac{m^2}{t^2}}\bigg)\bigg)+\sqrt{6} \alpha \big(72 m^3-12
m^2-t^3\big) \ln \bigg(\frac{6 m^2}{t^2}\bigg)\bigg)
\bigg(\frac{1}{m^2}\bigg[2 \sqrt{\frac{m^2}{t^2}}\\\nonumber
&\times& \bigg(\xi  t^7 \big(\sqrt{6} \alpha +9 c\big)
\sqrt{\frac{m^2}{t^2}}+m^2 t^4 \bigg(36 c \xi
\sqrt{\frac{m^2}{t^2}}+t^{3 m+1} \bigg(\sqrt{6} \alpha ^2+12 \alpha
c\\\nonumber &+&15 \beta  c \sqrt{\frac{m^2}{t^2}}+\sqrt{6} \alpha
\beta \sqrt{\frac{m^2}{t^2}}\bigg)\bigg)+36  t^{3 m+2}
\bigg(\sqrt{6} \alpha ^2-6 \alpha  c-12 \beta  c
\sqrt{\frac{m^2}{t^2}}\\\nonumber &+&5 \sqrt{6} \alpha  \beta
\sqrt{\frac{m^2}{t^2}}\bigg)m^5-12 m^4 t^{3 m+2} \bigg(\sqrt{6}
\alpha ^2-6 \alpha  c-9 \beta  c \sqrt{\frac{m^2}{t^2}}+4 \sqrt{6}
\alpha \beta \\\nonumber &\times&\sqrt{\frac{m^2}{t^2}}\bigg)+216
\sqrt{6} \beta ^2 m^7 t^{3 m}-48 \sqrt{6} \beta ^2 m^6 t^{3
m}\bigg)\bigg]+\sqrt{6} \alpha \ln \bigg(\frac{6 m^2}{t^2}\bigg)
\\\nonumber
&\times&\bigg(144 \beta  m^5 t^{3 m}-36 \beta m^4 t^{3 m}-m^2 t^2
\bigg(24 \alpha \sqrt{\frac{m^2}{t^2}} t^{3 m}+5 \beta  t^{3 m+1}+12
\xi \bigg)\\\nonumber &-&t^5 \bigg(4 \alpha \sqrt{\frac{m^2}{t^2}}
t^{3 m}+3 \xi \bigg)+\frac{72 \alpha  m^5 t^{3
m}}{\sqrt{\frac{m^2}{t^2}}}\bigg)\bigg)\bigg\}\bigg\{31104 m^4
\bigg(\alpha +2 \beta \sqrt{\frac{m^2}{t^2}}\bigg) \\\label{2ee}
&\times&\sqrt{\frac{m^2}{t^2}}\bigg(\beta  m^2 t^{3 m}+\alpha
\sqrt{\frac{m^2}{t^2}} t^{3 m+2}+\xi t^2\bigg)^2\bigg\}^{-1}.
\end{eqnarray}
Figure \textbf{12} demonstrates that the squared speed of sound is
positive for all considered values of $m$ which indicates the model
is stable.
\begin{figure}[H]\center
\epsfig{file=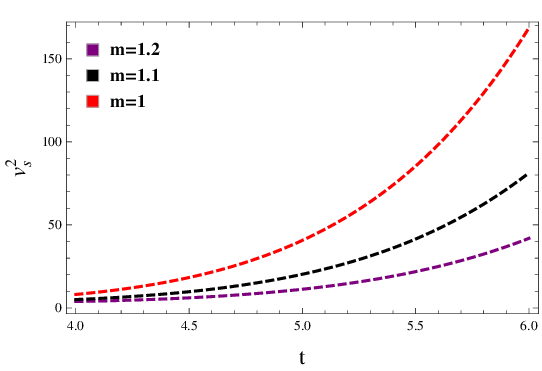,width=.6\linewidth}\caption{Graph of
$\nu_s^{2}$ versus $t$.}
\end{figure}

\section{Conclusions}

The main goal of this manuscript is to achieve a comprehensive
understanding of the cosmic behavior and consequences of DE models
in the framework of cosmic evolution. Initially, we have
reconstructed the DE $f(Q)$ gravity models using the correspondence
scheme which serves as a fundamental approach in constructing DE
models. We have employed the FRW model with a power-law scale factor
to investigate the non-interacting scenario. Our analysis has delved
into the exploration of the evolution trajectories of the EoS
parameter. We have extensively investigated the
$(\omega_{D}-\omega^{\prime}_{D})$ and $(r-s)$-planes which offer
valuable insights into the phase space dynamics of the GDE and GGDE
models. The main findings are summarized as follows.
\begin{itemize}
\item
For various values of $m$, the resulting $f(Q)$ gravity of both DE
models indicate an increasing trend over time, indicating that the
reconstructed models are realistic (Figures \textbf{1} and
\textbf{7}).
\item
Both DE models show an increasing trend for energy density, while
the pressure shows consistently negative characteristic for all
values of $m$. This behavior is consistent with the typical features
of DE (Figures \textbf{2} and \textbf{8}).
\item
The EoS indicates the existence of phantom field DE in both models.
This parameter becomes more negative below -1 which is consistent
with the current understanding of accelerated cosmic expansion
(Figures \textbf{3} and \textbf{9}).
\item
The freezing area for all values of $m$ is depicted by the
evolutionary behavior in the
($\omega_{D}$-$\omega^{\prime}_{D}$)-plane. This observation
provides confirmation that the non-interacting GGD and GGDE $f(Q)$
gravity models suggest a more rapid expansion of the universe
(Figures \textbf{4} and \textbf{10}).
\item
For both models, the $(r-s)$-plane depicts the Chaplygin gas model
for various values of $m$ (Figures \textbf{5} and \textbf{11}).
\item
It is found that $\nu_{s}^{2}<0$ for the GDE $f(Q)$ gravity model
indicating the instability (Figure \textbf{6}) but $\nu_{s}^{2}>0$
for the GGDE $f(Q)$ gravity model, suggesting stability across all
values of $m$ (Figure \textbf{12}).
\end{itemize}

We have observed that the findings align with the observational data
\cite{53} as indicated
\begin{eqnarray}\nonumber
\omega_{D}&=&-1.023^{+0.091}_{-0.096}\quad(\text{Planck
TT+LowP+ext}),\\\nonumber
\omega_{D}&=&-1.006^{+0.085}_{-0.091}\quad(\text{Planck
TT+LowP+lensing+ext}),\\\nonumber
\omega_{D}&=&-1.019^{+0.075}_{-0.080}\quad (\text{Planck TT, TE,
EE+LowP+ext}).
\end{eqnarray}
These values have been determined reaching a confidence level of
85\%. Notably, the EoS in the GGDE model exhibits a closer alignment
with observational data as compared to that of the GDE model. Our
results are consistent with the most recent theoretical
observational findings \cite{25a,45,53}.

We have noted that the results obtained from both the reconstruction
scheme in non-interacting ghost dark energy GDE and generalized
ghost dark energy GGDE align with modern observational data, as well
as with the studies conducted by \cite{16}-\cite{18}. Additionally,
Malekjani \cite{2ab} used the GGDE model to explore the current
cosmic expansion, both with and without interaction terms, to study
its cosmological evolution. Zubair and Abbas \cite{3ab} developed
the $f(R,\mathbb{T})$ gravity using the GDE model in a flat FRW
universe and found that their model includes both phantom and
quintessential phases. Myrzakulov et al. \cite{4ab} examined the
connection between $f(Q)$ and the GDE model, finding that the EoS
parameter crosses the phantom divide line. Our results align with
these findings. While this approach captures certain elements of the
early universe dynamics, it does not take into account
non-metricity. By including $Q$ in the model, we address both early
and late-time cosmic acceleration, providing a more complete picture
of the universe evolution. Comparing the results of the $f(Q)$
gravity model with these other theories, we show that this method
offers an observationally consistent model of the universe
accelerated expansion, contributing to a better understanding of DE
and cosmic evolution.

\section*{Appendix A: Calculation of $r$ and $s$ in GDE}
\renewcommand{\theequation}{A\arabic{equation}}
\setcounter{equation}{0}
\begin{eqnarray}\nonumber
r&=&\frac{1}{768 m^{10} \bigg(6 \sqrt{\frac{m^2}{t^2}} \alpha  \xi
t^{3 m}+6\bigg)^4}\bigg[24 \sqrt{6} t^{3 m+5} \xi
\bigg(\sqrt{\frac{m^2}{t^2}} \alpha  \xi  t^{3 m}+1\bigg)
\bigg(2\bigg(-c
\\\nonumber
&\times&864m^5\alpha ^2\xi ^2  t^{6 m}-24 m^4 \alpha ^2 \xi ^2
\big(\sqrt{6} \alpha -12 c\big) t^{6 m}+\bigg(18 \sqrt{6}
\sqrt{\frac{m^2}{t^2}} \alpha ^2 \xi  t^{3 m}\\\nonumber &+&111
\sqrt{\frac{m^2}{t^2}} \alpha  \xi  c t^{3 m}+8 \sqrt{6} \alpha +45
c\bigg) t^5- \bigg(3 \sqrt{\frac{m^2}{t^2}} \alpha \xi  t^{3
m}+1\bigg) \big(\sqrt{6} \alpha +6 c\big) \\\nonumber &\times&36 m^3
t^2+2 m^2 \bigg(6 \alpha ^2 \xi  \bigg(6 \xi c t^{3 m+1}+\sqrt{6}
\sqrt{\frac{m^2}{t^2}}\bigg) t^{3 m}+5 \sqrt{6} \alpha ^3 \xi ^2
t^{6 m+1}+54 c\\\nonumber &+&6 \alpha  \bigg(27
\sqrt{\frac{m^2}{t^2}} \xi  c t^{3 m}+\sqrt{6}\bigg)\bigg)
t^2\bigg)+\sqrt{6} \alpha \bigg(288 m^5 \alpha ^2 \xi ^2 t^{6 m}-96
m^4 \alpha ^2 \xi ^2 t^{6 m}\\\nonumber &-&\bigg(37
\sqrt{\frac{m^2}{t^2}} \alpha \xi  t^{3 m}+15\bigg) t^5+72 m^3
\bigg(3 \sqrt{\frac{m^2}{t^2}} \alpha  \xi  t^{3 m}+1\bigg) t^2-12
m^2 \bigg(9\alpha  \xi \\\nonumber &\times&\sqrt{\frac{m^2}{t^2}}
t^{3 m}+2 \alpha ^2 \xi ^2 t^{6 m+1}+3\bigg) t^2\bigg) \ln
\bigg(\frac{6 m^2}{t^2}\bigg)\bigg)
\bigg(\frac{m^2}{t^2}\bigg)^{3/2}+\bigg(\sqrt{6} t^{3 m} \alpha \xi
\\\nonumber &\times&\bigg(24 m^4 \alpha
\xi  t^{3 m}-72 m^5 \alpha  \xi  t^{3 m}+3 \sqrt{\frac{m^2}{t^2}}
t^5+ \bigg(\alpha  \xi  t^{3 m+1}+3 \sqrt{\frac{m^2}{t^2}}\bigg)t^24
m^2 \bigg)\\\nonumber &\times& \ln \bigg(\frac{6 m^2}{t^2}\bigg)-2
\bigg(-\frac{36 \sqrt{6} m^5 \alpha \xi  t^{3
m}}{\sqrt{\frac{m^2}{t^2}}}-12 m^4 \alpha  \xi ^2 \bigg(\sqrt{6}
\alpha -6 c\bigg) t^{6 m}+m^2 \xi  \\\nonumber &\times&\bigg(\alpha
\xi \bigg(\sqrt{6} \alpha +12 c\bigg) t^{3 m+1}+36
\sqrt{\frac{m^2}{t^2}} c\bigg) t^{3 m+2}+\sqrt{\frac{m^2}{t^2}} \xi
\big(\sqrt{6} \alpha +9 c\big) t^{3 m+5}\\\nonumber &+&144 \sqrt{6}
m^7 \alpha ^2 \xi ^2 t^{6 m-2}+72 m^5 \bigg(4 \sqrt{6}
\sqrt{\frac{m^2}{t^2}} \alpha  \xi  t^{3 m}-3 \alpha  \xi ^2 c t^{6
m}+2 \sqrt{6}\bigg)\bigg)\bigg)^2\\\nonumber &+&288 \sqrt{6} m^5
\bigg(\sqrt{\frac{m^2}{t^2}} \alpha  \xi  t^{3 m}+1\bigg)^2 \bigg(2
\bigg(-\frac{36 \sqrt{6} m^5 \alpha  \xi  t^{3
m}}{\sqrt{\frac{m^2}{t^2}}}-12 m^4 \alpha  \xi ^2 t^{6 m}\\\nonumber
&\times&\big(\sqrt{6} \alpha -6 c\big)+m^2 \xi \bigg(\alpha  \xi
\big(\sqrt{6} \alpha +12 c\big) t^{3 m+1}+36 \sqrt{\frac{m^2}{t^2}}
c\bigg) t^{3 m+2}+\sqrt{\frac{m^2}{t^2}}\\\nonumber &\times& \xi
\bigg(\sqrt{6} \alpha +9 c\bigg) t^{3 m+5}+144 \sqrt{6} m^7 \alpha
^2 \xi ^2 t^{6 m-2}+72 m^5 \bigg(4 \sqrt{6} \sqrt{\frac{m^2}{t^2}}
\alpha  \xi t^{3 m}\\\nonumber &-&3 \alpha  \xi ^2 c t^{6 m}+2
\sqrt{6}\bigg)\bigg)-\sqrt{6} t^{3 m}   \xi \bigg(24 m^4 \alpha \xi
t^{3 m}-72 m^5 \alpha  \xi  t^{3 m}+3 \sqrt{\frac{m^2}{t^2}}
t^5\\\nonumber &+&4 m^2 \bigg(\alpha \xi t^{3 m+1}+3
\sqrt{\frac{m^2}{t^2}}\bigg) t^2\bigg) \alpha\ln \bigg(\frac{6
m^2}{t^2}\bigg)\bigg)\bigg],
\\\nonumber
s&=&-\bigg\{16 \sqrt{\frac{2}{3}} m
\bigg(\frac{m^2}{t^2}\bigg)^{3/2} t^{2-3 m} \bigg(6
\sqrt{\frac{m^2}{t^2}} \alpha  \xi  t^{3 m}+6\bigg)^2
\bigg(\bigg[\bigg(\sqrt{\frac{m^2}{t^2}} \alpha  \xi t^{3 m}+1\bigg)
\\\nonumber &\times&24 \sqrt{6} t^{3 m+5} \xi \bigg(2 \bigg(-864 m^5 \alpha ^2 \xi ^2 c
t^{6 m}-24 m^4 \alpha ^2 \xi ^2 \big(\sqrt{6} \alpha -12 c\big) t^{6
m}\bigg(45 c\\\nonumber &+&18 \sqrt{6} \sqrt{\frac{m^2}{t^2}} \alpha
^2 \xi  t^{3 m}+111 \sqrt{\frac{m^2}{t^2}} \alpha  \xi  c t^{3 m}+8
\sqrt{6} \alpha \bigg) t^5-\big(\sqrt{6} \alpha +6 c\big)36 m^3
\\\nonumber &\times&\bigg(3 \sqrt{\frac{m^2}{t^2}} \alpha \xi  t^{3 m}+1\bigg)  t^2+2
m^2 \bigg(6 \alpha ^2 \xi \bigg(6 \xi c t^{3 m+1}+\sqrt{6}
\sqrt{\frac{m^2}{t^2}}\bigg) t^{3 m}+54 c\\\nonumber &+&5 \sqrt{6}
\alpha ^3 \xi ^2 t^{6 m+1}+6 \alpha  \bigg(27 \sqrt{\frac{m^2}{t^2}}
\xi  c t^{3 m}+\sqrt{6}\bigg)\bigg) t^2\bigg)+ \bigg(288 m^5 \alpha
^2 \xi ^2 t^{6 m}\\\nonumber &-&96 m^4 \alpha ^2 \xi ^2 t^{6
m}-\bigg(37 \sqrt{\frac{m^2}{t^2}} \alpha \xi  t^{3 m}+15\bigg)
t^5+72 m^3 \bigg(3 \sqrt{\frac{m^2}{t^2}} \alpha  \xi  t^{3
m}+1\bigg) t^2\\\nonumber &-&12 m^2 \bigg(9 \sqrt{\frac{m^2}{t^2}}
\alpha  \xi t^{3 m}+2 \alpha ^2 \xi ^2 t^{6 m+1}+3\bigg)
t^2\bigg)\sqrt{6} \alpha \ln \bigg(\frac{6 m^2}{t^2}\bigg)\bigg)
\bigg(\frac{m^2}{t^2}\bigg)^{3/2}\\\nonumber &+&\bigg(\sqrt{6} t^{3
m} \alpha \xi \bigg(-72 m^5 \alpha  \xi  t^{3 m}+24 m^4 \alpha  \xi
t^{3 m}+3 \sqrt{\frac{m^2}{t^2}} t^5+4 m^2 \bigg(\alpha  \xi  t^{3
m+1}\\\nonumber &+&3 \sqrt{\frac{m^2}{t^2}}\bigg) t^2\bigg) \ln
\bigg(\frac{6 m^2}{t^2}\bigg)-2 \bigg(-\frac{36 \sqrt{6} m^5 \alpha
\xi  t^{3 m}}{\sqrt{\frac{m^2}{t^2}}}-12 m^4 \alpha  \xi ^2
\big(\sqrt{6} \alpha -6 c\big) \\\nonumber &\times&t^{6 m}+m^2 \xi
\bigg(\alpha  \xi \big(\sqrt{6} \alpha +12 c\big) t^{3 m+1}+36
\sqrt{\frac{m^2}{t^2}} c\bigg)t^{3 m+2}+\big(\sqrt{6} \alpha +9
c\big) \xi\\\nonumber &\times&\sqrt{\frac{m^2}{t^2}} t^{3 m+5}+144
\sqrt{6} m^7 \alpha ^2 \xi ^2 t^{6 m-2}+72 m^5 \bigg(4 \sqrt{6}
\sqrt{\frac{m^2}{t^2}} \alpha  \xi  t^{3 m}+2 \sqrt{6}\\\nonumber
&-&3 \alpha  \xi ^2 c t^{6 m}\bigg)\bigg)\bigg)^2+288 \sqrt{6} m^5
\bigg(\sqrt{\frac{m^2}{t^2}} \alpha  \xi  t^{3 m}+1\bigg)^2 \bigg(2
\bigg(-\frac{36 \sqrt{6} m^5 \alpha  \xi  t^{3
m}}{\sqrt{\frac{m^2}{t^2}}}\\\nonumber &-&12 m^4 \alpha  \xi ^2
\big(\sqrt{6} \alpha -6 c\big) t^{6 m}+m^2 \xi  \bigg(\alpha
\big(\sqrt{6} \alpha +12 c\big) t^{3 m+1}+36 \sqrt{\frac{m^2}{t^2}}
c\bigg) t^{3 m+2}\\\nonumber &+&\sqrt{\frac{m^2}{t^2}} \xi
\bigg(\sqrt{6} \alpha +9 c\bigg) t^{3 m+5}+144 \sqrt{6} m^7 \alpha
^2 \xi ^2 t^{6 m-2}+ \bigg(4 \sqrt{6} \sqrt{\frac{m^2}{t^2}} \alpha
\xi  t^{3 m}\\\nonumber &-&3 \alpha  \xi ^2 c t^{6 m}+2
\sqrt{6}\bigg)72 m^5\bigg)-\sqrt{6} t^{3 m} \alpha  \xi \bigg(24 m^4
\alpha  \xi t^{3 m}-72 m^5 \alpha  \xi t^{3 m}+t^5\\\nonumber
&\times&3 \sqrt{\frac{m^2}{t^2}} +4 m^2 \bigg(\alpha  \xi  t^{3
m+1}+3 \sqrt{\frac{m^2}{t^2}}\bigg) t^2\bigg) \ln \bigg(\frac{6
m^2}{t^2}\bigg)\bigg)\bigg]\bigg[768 m^{10}
\bigg(6\sqrt{\frac{m^2}{t^2}}\\\nonumber &\times& \alpha \xi t^{3
m}+6\bigg)^4 \bigg]^{-1}-1\bigg)\bigg\}\bigg\{\xi \bigg(2 \bigg(36
\alpha \bigg(6 \sqrt{\frac{m^2}{t^2}} \xi  c t^{3 m}+\sqrt{6}\bigg)
m^3+\bigg(\sqrt{\frac{m^2}{t^2}}\\\nonumber &\times& t^{3 m} \alpha
\xi \big(\sqrt{6} \alpha -6 c\big)-3 c\bigg) 12 m^2-t^3
\bigg(\sqrt{6} \sqrt{\frac{m^2}{t^2}} \alpha ^2 \xi  t^{3 m}+9
c+\alpha  \bigg(\sqrt{\frac{m^2}{t^2}}\\\nonumber &\times& 12\xi  c
t^{3 m}+\sqrt{6}\bigg)\bigg)\bigg)-\sqrt{6} \alpha  \bigg(72 m^3
\sqrt{\frac{m^2}{t^2}} \alpha  \xi  t^{3 m}-\bigg(4
\sqrt{\frac{m^2}{t^2}} \alpha  \xi  t^{3 m}+3\bigg) t^3\\\nonumber
&-&12 m^2 \bigg(2 \sqrt{\frac{m^2}{t^2}} \alpha  \xi  t^{3
m}+1\bigg)\bigg) \ln \bigg(\frac{6
m^2}{t^2}\bigg)\bigg)\bigg\}^{-1}.
\end{eqnarray}

\section*{Appendix B: Calculation of $r$ and $s$ in GGDE}
\renewcommand{\theequation}{B\arabic{equation}}
\setcounter{equation}{0}
\begin{eqnarray}\nonumber
r&=&\bigg\{24 \sqrt{6} t^{3 m+7} \bigg(m^2 \beta  t^{3
m}+\sqrt{\frac{m^2}{t^2}} \alpha  t^{3 m+2}+\xi  t^2\bigg) \bigg(2
\bigg(72 m^7 \beta ^2 \bigg(9 \sqrt{6} \alpha +8
\\\nonumber&\times&\sqrt{6} \sqrt{\frac{m^2}{t^2}} \beta -27 c\bigg) t^{6
m}-12 m^6 \beta ^2 \bigg(25 \sqrt{6} \alpha +16 \sqrt{6}
\sqrt{\frac{m^2}{t^2}} \beta -45 c\bigg) t^{6 m}\\\nonumber&+&36 m^5
\bigg(6 t^{3 m} \alpha  \bigg(\sqrt{6} \sqrt{\frac{m^2}{t^2}} \alpha
\beta -11 \sqrt{\frac{m^2}{t^2}} c \beta -4 \alpha  c\bigg)-\beta
\xi \bigg(7 \sqrt{6} \alpha +12 c\\\nonumber&+&8 \sqrt{6}
\sqrt{\frac{m^2}{t^2}} \beta \bigg)\bigg) t^{3 m+2}+3 m^4 \bigg(-4
\alpha  \bigg(2 \sqrt{6} \alpha ^2+13 \sqrt{6}
\sqrt{\frac{m^2}{t^2}} \beta  \alpha -24 c \alpha \\\nonumber&-&63
\sqrt{\frac{m^2}{t^2}} \beta  c\bigg) t^{3 m}+\beta ^2 \big(4
\sqrt{6} \alpha +35 c\big) t^{3 m+1}+120 \beta  \xi c\bigg) t^{3
m+2}+\xi  \bigg(3 \sqrt{\frac{m^2}{t^2}} \\\nonumber&\times&\alpha
\big(6 \sqrt{6} \alpha +37 c\big) t^{3 m}+\xi  \big(8 \sqrt{6}
\alpha +45 c\big)\bigg) t^7-36 m^3 \xi  \bigg(3
\sqrt{\frac{m^2}{t^2}} \alpha t^{3 m}+\xi \bigg)t^4
\\\nonumber&\times&\big(\sqrt{6} \alpha +6 c\big) +m^2 \bigg(12
\sqrt{\frac{m^2}{t^2}} \alpha \xi \big(\sqrt{6} \alpha +27 c\big)
t^{3 m}+2 \beta  \xi \bigg(10 \sqrt{6} \alpha +63
c\bigg)\\\nonumber&\times& t^{3 m+1}+\alpha  \bigg(10 \sqrt{6}
\alpha ^2+22 \sqrt{6} \sqrt{\frac{m^2}{t^2}} \beta  \alpha +72 c
\alpha +171 \sqrt{\frac{m^2}{t^2}} \beta  c\bigg) t^{6 m+1}+\xi
^2\\\nonumber&\times& 12 \big(\sqrt{6} \alpha +9 c\big)\bigg)
t^4\bigg)+\sqrt{6} \alpha \bigg(648 m^7 \beta ^2 t^{6 m}-180 m^6
\beta ^2 t^{6 m}+72 m^5 \bigg(\bigg(4 \alpha \\\nonumber&+&11
\sqrt{\frac{m^2}{t^2}} \beta \bigg) \alpha t^{3 m}+2 \beta  \xi
\bigg) t^{3 m+2}-m^4 \bigg(12 \alpha \bigg(8 \alpha +21
\sqrt{\frac{m^2}{t^2}} \beta \bigg) t^{3 m}+35 \beta
^2\\\nonumber&\times& t^{3 m+1}+120 \beta  \xi \bigg) t^{3 m+2}-\xi
\bigg(37 \sqrt{\frac{m^2}{t^2}} \alpha  t^{3 m}+15 \xi \bigg) t^7+
\bigg(3 \sqrt{\frac{m^2}{t^2}} \alpha t^{3 m}+\xi \bigg)
\\\nonumber&\times&72 m^3 \xi t^4-3 m^2 \bigg(36
\sqrt{\frac{m^2}{t^2}} \alpha \xi t^{3 m}+14 \beta  \xi t^{3
m+1}+\alpha  \bigg(8 \alpha +19 \sqrt{\frac{m^2}{t^2}} \beta \bigg)
t^{6 m+1}\\\nonumber&+&12 \xi ^2\bigg) t^4\bigg) \ln \bigg(\frac{6
m^2}{t^2}\bigg)\bigg) \bigg(\frac{m^2}{t^2}\bigg)^{\frac{3}{2}}+
\sqrt{6} m^5 \bigg(m^2 \beta t^{3 m}+\sqrt{\frac{m^2}{t^2}} \alpha
t^{3 m+2}+\xi t^2\bigg)^2 \\\nonumber&\times&\bigg(2 \bigg(144
\sqrt{6} m^7 \beta \bigg(2 \alpha +\sqrt{\frac{m^2}{t^2}} \beta
\bigg) t^{6 m}+72 m^5 \bigg(\bigg(2 \sqrt{6} \sqrt{\frac{m^2}{t^2}}
\alpha ^2+\sqrt{6} \beta  \alpha \\\nonumber&+&2 \sqrt{6}
\sqrt{\frac{m^2}{t^2}} \beta ^2-6 \beta c\bigg) t^{3 m}+4 \sqrt{6}
\bigg(\alpha +\sqrt{\frac{m^2}{t^2}} \beta \bigg) \xi \bigg) t^{3
m+2}+m^2 \bigg(36 \xi c\\\nonumber&-&12 \sqrt{\frac{m^2}{t^2}}
\alpha \big(\sqrt{6} \alpha -6 c\big) t^{3 m}+\beta \big(\sqrt{6}
\alpha +15 c\big) t^{3 m+1}\bigg) t^{3
m+4}+\bigg(\sqrt{\frac{m^2}{t^2}} \alpha t^{3 m}
\\\nonumber&\times&\big(\sqrt{6} \alpha +12 c\big)+\xi
\big(\sqrt{6} \alpha +9 c\big)\bigg) t^{3 m+7}-12  \bigg(4 \sqrt{6}
\alpha +4 \sqrt{6} \sqrt{\frac{m^2}{t^2}} \beta -9
c\bigg)\\\nonumber&\times& m^4 \beta t^{6 m+2}-\bigg(\sqrt{6}
\bigg(\alpha +2 \sqrt{\frac{m^2}{t^2}} \beta \bigg) \xi  t^{3 m}+6
\sqrt{\frac{m^2}{t^2}} \alpha  c t^{6 m}-4 \sqrt{6}
\sqrt{\frac{m^2}{t^2}} \xi ^2\bigg)\\\nonumber&\times&36 m^3
t^4\bigg)-\sqrt{6} \bigg(36 m^4 \beta  t^{3 m}-144 m^5 \beta  t^{3
m}-\frac{72 m^5 \alpha  t^{3 m}}{\sqrt{\frac{m^2}{t^2}}}+\bigg(4
\sqrt{\frac{m^2}{t^2}} \alpha t^{3 m}\\\nonumber&+&3 \xi \bigg)
t^5+m^2 \bigg(24 \sqrt{\frac{m^2}{t^2}} \alpha t^{3 m}+5 \beta  t^{3
m+1}+12 \xi \bigg) t^2\bigg) t^{3 m+2} \alpha \ln \bigg(\frac{6
m^2}{t^2}\bigg)\bigg)\\\nonumber&\times& \sqrt{\frac{m^2}{t^2}}+m^2
\bigg(\sqrt{6} t^{3 m+2} \alpha \bigg(36 m^4 \beta  t^{3 m}-144 m^5
\beta  t^{3 m}-\frac{72 m^5 \alpha  t^{3
m}}{\sqrt{\frac{m^2}{t^2}}}+t^5\\\nonumber&\times& \bigg(4
\sqrt{\frac{m^2}{t^2}} \alpha t^{3 m}+3 \xi \bigg) -2 \bigg(144
\sqrt{6} m^7 \beta \bigg(2 \alpha +\sqrt{\frac{m^2}{t^2}} \beta
\bigg)  \bigg(24 \sqrt{\frac{m^2}{t^2}} \alpha t^{3
m}\\\nonumber&+&5 \beta  t^{3 m+1}+12 \xi \bigg) t^{6 m}m^2t^2\bigg)
\ln \bigg(\frac{6 m^2}{t^2}\bigg)+72 m^5 \bigg(\bigg(2 \sqrt{6}
\sqrt{\frac{m^2}{t^2}} \alpha ^2 -6 \beta c\\\nonumber&+&\sqrt{6}
\beta  \alpha+2 \sqrt{6} \sqrt{\frac{m^2}{t^2}} \beta ^2\bigg) t^{3
m}+4 \sqrt{6} \bigg(\alpha +\sqrt{\frac{m^2}{t^2}} \beta \bigg) \xi
\bigg) t^{3 m+2}+m^2 t^{3 m+4}\\\nonumber&\times&\bigg(36 \xi c-12
\sqrt{\frac{m^2}{t^2}} \alpha \big(\sqrt{6} \alpha -6 c\big) t^{3
m}+\beta \big(\sqrt{6} \alpha +15 c\big) t^{3 m+1}\bigg)
+\bigg(\sqrt{\frac{m^2}{t^2}} \\\nonumber&\times&\alpha
\big(\sqrt{6} \alpha +12 c\big) t^{3 m}+\xi  \big(\sqrt{6} \alpha +9
c\big)\bigg) t^{3 m+7}-12 \bigg(4 \sqrt{6} \alpha +4 \sqrt{6}
\sqrt{\frac{m^2}{t^2}} \beta \\\nonumber&-&9 c\bigg)m^4 \beta t^{6
m+2}-36 m^3 \bigg(\sqrt{6} \bigg(\alpha +2 \sqrt{\frac{m^2}{t^2}}
\beta \bigg) \xi  t^{3 m}+6 \sqrt{\frac{m^2}{t^2}} \alpha  c t^{6
m}-4\\\nonumber&\times& \sqrt{6} \sqrt{\frac{m^2}{t^2}} \xi ^2\bigg)
t^4\bigg)\bigg)^2\bigg\}\bigg\{768 m^{10} t^6 \bigg(\bigg(\frac{6
\beta m^2}{t^2}+6 \sqrt{\frac{m^2}{t^2}} \alpha \bigg) t^{3 m}+6 \xi
\bigg)^4\bigg\}^{-1},\\\nonumber s&=&\bigg\{16 \sqrt{\frac{2}{3}} m
\bigg(\frac{m^2}{t^2}\bigg)^{3/2} t^{4-3 m} \bigg(\bigg(\frac{6
\beta  m^2}{t^2}+6 \sqrt{\frac{m^2}{t^2}} \alpha \bigg) t^{3 m}+6
\xi \bigg)^2 \bigg(\bigg\{24 \sqrt{6}\\\nonumber&\times& t^{3 m+7}
\bigg(m^2 \beta t^{3 m}+\sqrt{\frac{m^2}{t^2}} \alpha  t^{3 m+2}+\xi
t^2\bigg) \bigg(2 \bigg(72  \bigg(9 \sqrt{6} \alpha +8 \sqrt{6}
\sqrt{\frac{m^2}{t^2}} \beta \\\nonumber&-&27 c\bigg) m^7 \beta
^2t^{6 m}-12 m^6 \beta ^2 \bigg(25 \sqrt{6} \alpha +16 \sqrt{6}
\sqrt{\frac{m^2}{t^2}} \beta -45 c\bigg) t^{6 m}+36 m^5
\\\nonumber&\times&\bigg(6 t^{3 m} \alpha  \bigg(\sqrt{6}
\sqrt{\frac{m^2}{t^2}} \alpha  \beta -11 \sqrt{\frac{m^2}{t^2}} c
\beta -4 \alpha  c\bigg)-\beta \bigg(7 \sqrt{6} \alpha +8 \sqrt{6}
\sqrt{\frac{m^2}{t^2}} \beta \\\nonumber&+&12 c\bigg)\xi\bigg) t^{3
m+2}+3 m^4 \bigg(-4 \alpha  \bigg(2 \sqrt{6} \alpha ^2+13 \sqrt{6}
\sqrt{\frac{m^2}{t^2}} \beta  \alpha -24 c \alpha -63\beta
\\\nonumber&\times&\sqrt{\frac{m^2}{t^2}}   c\bigg) t^{3 m}+\beta ^2
\bigg(4 \sqrt{6}\alpha +35 c\bigg) t^{3 m+1}+120 \beta  \xi c\bigg)
t^{3 m+2}+\xi \bigg(3 \sqrt{\frac{m^2}{t^2}}\\\nonumber&\times&
\alpha \big(6 \sqrt{6} \alpha +37 c\big) t^{3 m}+\xi  \big(8
\sqrt{6} \alpha +45 c\big)\bigg) t^7-36 m^3 \xi  \bigg(3
\sqrt{\frac{m^2}{t^2}} \alpha t^{3 m}+\xi \bigg)\\\nonumber&\times&
\big(\sqrt{6} \alpha +6 c\big) t^4+m^2 \bigg(12
\sqrt{\frac{m^2}{t^2}} \alpha \xi \big(\sqrt{6} \alpha +27 c\big)
t^{3 m}+2   \xi \big(10 \sqrt{6} \alpha +63 c\big)
\\\nonumber&\times&\beta t^{3 m+1}+\alpha  \bigg(10 \sqrt{6} \alpha ^2+22
\sqrt{6} \sqrt{\frac{m^2}{t^2}} \beta  \alpha +72 c \alpha +171
\sqrt{\frac{m^2}{t^2}} \beta  c\bigg) t^{6 m+1}\\\nonumber&+&12 \xi
^2 \bigg(\sqrt{6} \alpha +9 c\bigg)\bigg) t^4\bigg)+\sqrt{6} \alpha
\bigg(648 m^7 \beta ^2 t^{6 m}-180 m^6 \beta ^2 t^{6 m}+72 m^5
\\\nonumber&\times&\bigg(\alpha  \bigg(4 \alpha +11 \sqrt{\frac{m^2}{t^2}} \beta \bigg)
t^{3 m}+2 \beta  \xi \bigg) t^{3 m+2}-m^4 \bigg(12 \alpha \bigg(8
\alpha +21 \sqrt{\frac{m^2}{t^2}} \beta
\bigg)\\\nonumber&\times&t^{3 m}+35 \beta ^2 t^{3 m+1}+120 \beta \xi
\bigg) t^{3 m+2}-\xi  \bigg(37 \sqrt{\frac{m^2}{t^2}} \alpha t^{3
m}+15 \xi \bigg) t^7+72 m^3 \xi\\\nonumber&\times& \bigg(3
\sqrt{\frac{m^2}{t^2}} \alpha  t^{3 m}+\xi \bigg) t^4-3 m^2 \bigg(36
\sqrt{\frac{m^2}{t^2}} \alpha  \xi  t^{3 m}+14 \beta  \xi t^{3
m+1}+12 \xi ^2+\alpha \\\nonumber&\times&\bigg(8 \alpha +19
\sqrt{\frac{m^2}{t^2}} \beta \bigg) t^{6 m+1}\bigg) t^4\bigg) \ln
\bigg(\frac{6 m^2}{t^2}\bigg)\bigg)
\bigg(\frac{m^2}{t^2}\bigg)^{\frac{3}{2}}+288 \sqrt{6} m^5 \bigg(\xi
t^2\\\nonumber&+&m^2 \beta t^{3 m}+\sqrt{\frac{m^2}{t^2}} \alpha
t^{3 m+2}\bigg)^2 \bigg(2 \bigg(144 \sqrt{6} m^7 \beta \bigg(2
\alpha +\sqrt{\frac{m^2}{t^2}} \beta \bigg) t^{6 m}+72 m^5
\\\nonumber&\times&\bigg(\bigg(2 \sqrt{6} \sqrt{\frac{m^2}{t^2}} \alpha ^2+\sqrt{6}
\beta  \alpha +2 \sqrt{6} \sqrt{\frac{m^2}{t^2}} \beta ^2-6 \beta
c\bigg) t^{3 m}+ \bigg(\alpha +\sqrt{\frac{m^2}{t^2}} \beta
\bigg)\\\nonumber&\times&4 \sqrt{6} \xi \bigg) t^{3 m+2}+m^2
\bigg(-12 \sqrt{\frac{m^2}{t^2}} \alpha \big(\sqrt{6} \alpha -6
c\big) t^{3 m}+\beta \big(\sqrt{6} \alpha +15 c\big) t^{3
m+1}\\\nonumber&+&36 \xi c\bigg) t^{3
m+4}+\bigg(\sqrt{\frac{m^2}{t^2}} \alpha \big(\sqrt{6} \alpha +12
c\big) t^{3 m}+\xi  \big(\sqrt{6} \alpha +9 c\big)\bigg) t^{3
m+7}-12 \\\nonumber&\times&m^4 \beta  \bigg(4 \sqrt{6} \alpha +4
\sqrt{6} \sqrt{\frac{m^2}{t^2}} \beta -9 c\bigg) t^{6 m+2}-36 m^3
\bigg(\sqrt{6} \bigg(\alpha +2 \sqrt{\frac{m^2}{t^2}} \beta \bigg)
\xi \\\nonumber&\times&t^{3 m}+6 \sqrt{\frac{m^2}{t^2}} \alpha  c
t^{6 m}-4 \sqrt{6} \sqrt{\frac{m^2}{t^2}} \xi ^2\bigg)
t^4\bigg)-\sqrt{6} t^{3 m+2} \alpha  \bigg(-144 m^5 \beta  t^{3
m}\\\nonumber&+&36 m^4 \beta  t^{3 m}-\frac{72 m^5 \alpha  t^{3
m}}{\sqrt{\frac{m^2}{t^2}}}+\bigg(4 \sqrt{\frac{m^2}{t^2}} \alpha
t^{3 m}+3 \xi \bigg) t^5+m^2 \bigg(24 \sqrt{\frac{m^2}{t^2}} \alpha
t^{3 m}\\\nonumber&+&5 \beta  t^{3 m+1}+12 \xi \bigg) t^2\bigg) \ln
\bigg(\frac{6 m^2}{t^2}\bigg)\bigg) \sqrt{\frac{m^2}{t^2}}+m^2
\bigg(\sqrt{6} t^{3 m+2} \alpha \bigg(36 m^4 \beta  t^{3
m}\\\nonumber&-&144 m^5 \beta t^{3 m}-\frac{72 m^5 \alpha  t^{3
m}}{\sqrt{\frac{m^2}{t^2}}}+\bigg(4 \sqrt{\frac{m^2}{t^2}} \alpha
t^{3 m}+3 \xi \bigg) t^5+m^2 \bigg(24 \sqrt{\frac{m^2}{t^2}} \alpha
t^{3 m}\\\nonumber&+&5 \beta  t^{3 m+1}+12 \xi \bigg) t^2\bigg) \ln
\bigg(\frac{6 m^2}{t^2}\bigg)-2 \bigg(144 \sqrt{6} m^7 \beta \bigg(2
\alpha +\sqrt{\frac{m^2}{t^2}} \beta \bigg) t^{6 m}\\\nonumber&+&72
m^5 \bigg(\bigg(2 \sqrt{6} \sqrt{\frac{m^2}{t^2}} \alpha ^2+\sqrt{6}
\beta  \alpha +2 \sqrt{6} \sqrt{\frac{m^2}{t^2}} \beta ^2-6 \beta
c\bigg) t^{3 m}+4 \sqrt{6} \\\nonumber&\times&\bigg(\alpha
+\sqrt{\frac{m^2}{t^2}} \beta \bigg) \xi \bigg) t^{3 m+2}+m^2
\bigg(\beta  \big(\sqrt{6} \alpha +15 c\big) t^{3 m+1}-12
\sqrt{\frac{m^2}{t^2}} \alpha t^{3
m}\\\nonumber&\times&\big(\sqrt{6} \alpha -6 c\big) +36 \xi c\bigg)
t^{3 m+4}+\bigg(\sqrt{\frac{m^2}{t^2}} \alpha \big(\sqrt{6} \alpha
+12 c\big) t^{3 m}+\xi  \big(\sqrt{6} \alpha +9
c\big)\bigg)\\\nonumber&\times& t^{3 m+7}-12 m^4 \beta  \bigg(4
\sqrt{6} \alpha +4 \sqrt{6} \sqrt{\frac{m^2}{t^2}} \beta -9 c\bigg)
t^{6 m+2}- \bigg( \bigg(\alpha +2 \sqrt{\frac{m^2}{t^2}} \beta
\bigg) \\\nonumber&\times&\sqrt{6} \xi  t^{3 m}+6
\sqrt{\frac{m^2}{t^2}} \alpha  c t^{6 m}-4 \sqrt{6}
\sqrt{\frac{m^2}{t^2}} \xi ^2\bigg)36 m^3
t^4\bigg)\bigg)^2\bigg\}\bigg\{768  \bigg(\bigg(\frac{6 \beta
m^2}{t^2}\\\nonumber&+&6 \sqrt{\frac{m^2}{t^2}} \alpha \bigg) t^{3
m}+6 \xi \bigg)^4 m^{10}
t^6\bigg\}^{-1}-1\bigg)\bigg\}\bigg\{\bigg[2 \sqrt{\frac{m^2}{t^2}}
\bigg(t^7 \xi  \big(\sqrt{6} \alpha +9 c\big)
\sqrt{\frac{m^2}{t^2}}\\\nonumber&-&36 \sqrt{6} m t^6 \alpha  \xi
\bigg(\frac{m^2}{t^2}\bigg)^{3/2}+144 \sqrt{6} m^7 t^{3 m} \beta
^2-48 \sqrt{6} m^6 t^{3 m} \beta ^2-12 m^4 t^{3
m+2}\\\nonumber&\times& \bigg(\sqrt{6} \alpha ^2+4 \sqrt{6}
\sqrt{\frac{m^2}{t^2}} \beta \alpha -6 c \alpha -9
\sqrt{\frac{m^2}{t^2}} \beta  c\bigg)+m^2 t^4 \bigg(\bigg(\sqrt{6}
\alpha ^2+12 c \alpha\\\nonumber&+&\sqrt{6} \sqrt{\frac{m^2}{t^2}}
\beta \alpha +15 \sqrt{\frac{m^2}{t^2}} \beta  c\bigg) t^{3 m+1}+36
\sqrt{\frac{m^2}{t^2}} \xi  c\bigg)+72 m^5 t^2 \bigg(t^{3 m}
\bigg(\sqrt{6}\alpha \\\nonumber&\times&\sqrt{\frac{m^2}{t^2}} \beta
-6 \sqrt{\frac{m^2}{t^2}} c \beta -3 \alpha  c\bigg)-\sqrt{6} \beta
\xi \bigg)\bigg)\bigg]\frac{1}{m^2}+\sqrt{6} \alpha \bigg(\frac{72
m^5 \alpha t^{3 m}}{\sqrt{\frac{m^2}{t^2}}}+\beta
\\\nonumber&\times&144 m^5t^{3 m}-36 m^4 \beta t^{3 m}-\bigg(4 \sqrt{\frac{m^2}{t^2}} \alpha
t^{3 m}+3 \xi \bigg) t^5-m^2 \bigg(24 \sqrt{\frac{m^2}{t^2}} \alpha
t^{3 m} \\\nonumber&+& 5 \beta t^{3 m+1}+12 \xi \bigg) t^2\bigg) \ln
\bigg(\frac{6 m^2}{t^2}\bigg)\bigg\}^{-1}.
\end{eqnarray}
\textbf{Data Availability Statement:} No data was used for the
research described in this paper.

\end{document}